\documentstyle[aps,epsf]{revtex}
\def\be{\begin{equation}}
\def\ee{\end{equation}}
\def\bea{\begin{eqnarray}}
\def\eea{\end{eqnarray}}
\begin{document}  

\title{Gauge Model For Maximal Neutrino Mixing}

\author{ Rabindra N. Mohapatra$^a$ and Shmuel Nussinov$^b$}
\address{\it$^a$ Department of Physics, University of Maryland,
College Park, MD 20742, USA\\
$^b$ Department of Physics, Tel Aviv University, Tel Aviv, Israel }

\maketitle

\begin{abstract}
The recently announced Super-Kamiokande data on atmospheric neutrino
oscillation seems to require a maximal mixing between the $\nu_{\mu}$
and $\nu_{\tau}$ within the conventional three neutrino picture. It is then
tempting to suggest as has been done in literature, that the solar
neutrino deficit be also understood as resulting from a maximal mixing
between $\nu_e$ and $\nu_{\mu}$. In this letter, we propose a 
left-right symmetric extension of the standard model where permutation 
symmetry leads to one of the maximal mixing patterns in a technically 
natural manner. The double seesaw mechanism gives small 
Majorana masses for neutrinos needed to 
understand the atmospheric as well as the solar neutrino puzzles.
\end{abstract}

\hskip 7cm UMD-PP-98-132
 
\section{Introduction}\hspace{0.5cm} 

The announcement by the Super-Kamiokande collaboration\cite{sk} of 
evidence for neutrino oscillation (and hence nonzero neutrino mass) in
their atmospheric neutrino data is a major milestone in  
the search for new physics beyond the standard model.
An outstanding feature of these oscillations 
 is the maximal mixing between the $\nu_{\mu}$ and 
$\nu_{\tau}$ ($sin^22\theta_{\mu-\tau}\approx 0.7-1$) in sharp contrast with 
the mixing pattern in the quark 
sector. Also the inferred $\Delta m^2_{\mu-\tau}\sim 5\times 10^{-4}-
6\times 10^{-3}$ eV$^2$ is lower than most ``see-saw motivated'' 
extrapolations from $\Delta m^2_{e-\mu}$ values in the small or large 
angle MSW solutions to the solar neutrino problem\cite{solar}: 
 $\Delta m^2_{e-\mu}\simeq 3\times 10^{-6}-7\times 10^{-6}$eV$^2$ 
with $sin^22\theta \simeq 3-5\times 10^{-3}$ and 
 $\Delta m^2\simeq 10^{-5}-10^{-4}$ eV$^2$ with $sin^22\theta \simeq .8-1$.

The solar neutrino problem provided the first indication for neutrino 
oscillation and this evidence keeps building up. It can also be resolved 
by the maximal $\nu_e-\nu_{\mu}$ vacuum oscillation with fine tuned 
small mass difference $\Delta m^2_{e-\mu}\approx 10^{-10}$ eV$^2$. 
Maximal mixing with larger $\Delta m^2$ values yield an energy independent
suppression of all solar neutrinos\footnote{More precisely, the 
suppression factor in the radio chemical experiments is by 50\% whereas 
in the Super-Kamiokande it is 57\%.} (except when it is in the 
large angle MSW range mentioned above). While this does not resolve the 
solar neutrino problem at present, it does considerably ameliorate it. 

All the above suggests considering maximal ($\nu_e-\nu_{\mu}$) mixing 
alongside maximal ($\nu_{\mu}-\nu_{\tau}$) mixing\cite{gold}. 
Three specific ``bimaximal mixing'' patterns\cite{nussinov,fritzsch,other,gold}
having particularly simple forms are: 

\noindent{\it Case (A)\cite{nussinov}:}
\begin{eqnarray}
U_{\nu}=\frac{1}{\sqrt{3}} \left(\begin{array}{ccc}
1 & \omega & \omega^2\\
1 & \omega^2 &\omega\\
1 & 1 & 1 \end{array} \right)
\end{eqnarray}
where $\omega =e^{\frac{2\pi i}{3}}$; we will call this the symmetric mixing
pattern.

\noindent{\it Case (B)\cite{gold}:}
\begin{eqnarray}
U_{\nu}=\left(\begin{array}{ccc}
\frac{1}{\sqrt{2}} & -\frac{1}{\sqrt{2}} & 0\\
\frac{1}{2} &\frac{1}{2} &\frac{1}{\sqrt{2}} \\
\frac{1}{2} &\frac{1}{2} &-\frac{1}{\sqrt{2}}
\end{array}\right)
\end{eqnarray}
This has been called in the literature as bimaximal mixing\cite{gold}.

\noindent{\it Case (C)\cite{fritzsch}:}
\begin{eqnarray}
U_{\nu}=\left(\begin{array}{ccc}
\frac{1}{\sqrt{2}} & -\frac{1}{\sqrt{2}} & 0\\
\frac{1}{\sqrt{6}} &\frac{1}{\sqrt{6}} &-\frac{2}{\sqrt{6}} \\
\frac{1}{\sqrt{3}} &\frac{1}{\sqrt{3}} &\frac{1}{\sqrt{3}}
\end{array}\right)
\end{eqnarray}
We call this democratic mixing.

In the above equations, we have defined $U_{\nu}$ as follows:
\begin{eqnarray}
\left(\begin{array}{c} \nu_e\\ \nu_{\mu} \\ \nu_{\tau}\end{array}
\right)=~ U_{\nu}\left(\begin{array}{c} \nu_1 \\ \nu_2 \\ \nu_3 
\end{array} \right) 
\end{eqnarray}
with $\nu_{ e,\mu, \tau}$ being the weak eigenstates and $\nu_{1,2,3}$, 
the mass eigenstates.

Our goal is to explore possible extensions of the standard model that may 
provide a theoretical understanding of 
the maximal mixing patterns. Attempts to understand the pattern (A),
made in our previous paper\cite{nussinov} were largely unsuccessful.
Also the CHOOZ upper bound\cite{chooz} on $\nu_e-\nu_{x}$ oscillation with
$\Delta m^2_{13}\geq 10^{-3}$ eV$^2$, tends together with 
Super-Kamiokande data, to disfavor this possibility. There has been 
several recent attempts to derive the second pattern (B)\cite{sacha}. 
Here we will show by using an extension of 
the standard model, that it is possible to generate the pattern (C) 
in a consistent and natural way. Our
motivation is quite clear: if nature presents
us with such a neutrino mixing pattern, we must seek an extension of the 
standard model that can naturally lead to it. Hopefully a theory that 
naturally provides this pattern will have other testable predictions.

In Ref.\cite{fritzsch}, permutation symmetry was imposed on the charged 
lepton mass matrix and not on the neutrino mass matrices in order to
 motivate the pattern (C). No underlying theoretical justification was 
discussed for such a hypothesis. In the framework of gauge theories, such 
an assumption is hard to understand apriori since the charged leptons and 
the neutrinos are members of the same isodoublet of the standard model 
gauge group $SU(2)_L$ and therefore the permutation symmetry could 
lead to a similar
mass matrix for both the charged lepton sector as well as the neutrino 
sector. If that happens, the neutrino mixing 
matrix which is given by $U^{\dagger}_{ell} U_{\nu}$ could substantially
differ from (C). It is therefore important to investigate whether the above
mixing pattern arises in a complete theory. Also the 
putative mass pattern $\Delta m^2_{32}\gg \Delta m^2_{21}$ should be 
provided by the model rather than arbitrarily fixed.
 It is considerations such as these which motivate
us to add this brief note to the exploding literature on neutrino models.   

We find that by combining the permutation symmetry $S_3$ with a 
$Z_4\times Z_3\times Z_2$ symmetry in the left-right symmetric extension of 
the standard model, we can obtain the 
maximal mixing pattern (C) in a technically natural manner (i.e. without 
setting any parameters to zero by hand). 
In the limit of exact permutation
symmetry, all the neutrinos are degenerate as are the elctron and the muon.
As a result, the mixing angles can be rotated away. However, once one
admits permutation breaking terms to accomodate the electron muon mass
difference, the neutrino degeneracy is removed and the democratic 
form (pattern C) of the maximal mixing
pattern remains. In fact, the masses of $\nu_e$ and $\nu_{\mu}$ get 
related to the elctron and muon masses arising completely from 
radiative corrections. To
avoid arbitrary deviations from the maximal pattern, we assume that
the permutation symmetry (but not the $Z_4\times Z_3\times Z_2$) is softly 
broken.
This adds only small, finite, corrections to the mixing pattern and one 
obtains a complete and realistic gauge theoretic derivation of the maximal 
mixing pattern C.

\section{Permutation symmetry and a gauge theory of maximal mixing}

We consider a left-right symmetric extension of the standard model with the 
usual fermionic field content\cite{lr}. We omit the discussion of the 
quark sector for now. Denoting the leptons by $\psi_a\equiv (\nu_a, e_a)$,
the $\psi_{L,R}$ transform as 
the $SU(2)_{L,R}$ doublets respectively under the left-right gauge group 
$SU(2)_L\times SU(2)_R\times U(1)_{B-L}$. The subscript $a$ 
stands for the generations. We choose the following set of 
Higgs bosons to achieve the symmetry breaking: three sets of left and 
right doublets dnoted by $\chi_{a,L,R}$ ($a=1,2,3$); the $\chi_{aR}$ vev 
will break the $SU(2)_R\times U(1)_{B-L}$ gauge symmetry down to the
standard model $U(1)_Y$ group. We choose three bidoublets $\phi_{0,1,2}$ to 
break the electroweak $SU(2)_L\times U(1)_Y$ symmetry and give mass to 
the quarks and charged leptons as well as the Dirac mass for the neutrinos. 
In order to implement the double seesaw\cite{valle} mechanism for 
neutrino masses, we introduce three gauge singlet fermion fields, $\sigma_a$.

In order to get the desired pattern for lepton masses, we demand the 
theory to respect the symmetry $S_3\times Z_4\times Z_3\times Z_2$ for all 
dimension
four terms. We assume that all interactions of dimension four are invariant
under permutation of the three indices $a=1,2,3$ for the fields that 
carry the subscript $a$\cite{derman}. This symmetry will be softly broken by 
terms of dimension $\leq 3$. We assume that under left-right symmetry 
$\phi_0\leftrightarrow \phi^{\dagger}_0$ and 
$\phi_1\leftrightarrow\phi^{\dagger}_2$ 
The transformation of the various fields under symmetry $Z_4\times Z_3\times 
Z_2$ is given in Table I. The quark fields are assumed to be singlets 
under the above groups.

The Yukawa couplings invariant under the above symmetries are:
\begin{eqnarray}
{\cal L_Y}= h_0 \Sigma_a \bar{\psi}_{aL}\phi_0\psi_{aR} + h_1 
(\bar{\psi}_{1L}\phi_1\psi_{2R} +\bar{\psi}_{2L}\phi_1 
\psi_{3R}+\bar{\psi}_{3L}\phi_1\psi_{1R})\nonumber\\
+h_1(\bar{\psi}_{1R}\phi^{\dagger}_2\psi_{2L} 
+\bar{\psi}_{2R}\phi^{\dagger}_2
\psi_{3L}+\bar{\psi}_{3R}\phi^{\dagger}_2\psi_{1L}) h.c. 
\end{eqnarray}
It is then clear that after the $\phi_{0,1,2}$ acquire vev's, they will 
give Dirac mass to the charged leptons and the neutrinos.
To get the desired pattern of charged lepton masses and the Dirac mass for 
the neutrinos, we choose the vev pattern for the $\phi$'s as follows:
$<\phi_0>=\left(\begin{array}{cc} \kappa_0 & 0\\
0 & \kappa'_0\end{array}\right)$. On the other hand, for the fields 
$\phi_{1,2}$, we choose $<\phi_{1,2}>=\left(\begin{array}{cc} 0 & 0\\
0 & \kappa'_{1,2}\end{array}\right)$. Due to left-right symmetry, one
can assure that $\kappa'_1=\kappa'_2$. It is crucial that the the 
vev pattern for $\phi_{1,2}$ is stable since this is what distinguishes 
the neutrino sector from the charged lepton sector and leads to the 
maximal mixing pattern (C) of democratic type for the neutrinos. It is 
important 
for this that there be no tadpole terms involving the $11$ components of the 
$\phi_{1,2}$ fields. This is verified by making the 
observation that all the $\phi_\alpha$ have same $Z_4$ quantum number; as a 
result terms like $Tr(\tilde{\phi}_\alpha\phi_\alpha)$ which could generate 
the tadpoles are not present in the potential. 

The above vev pattern has the consequence that all elements of the charged 
lepton mass matrix are nonzero whereas the Dirac mass matrix for the 
neutrinos is diagonal. To see the resulting mixing matrix, let us write the 
charged lepton mass matrix:
 \begin{eqnarray} 
M_{\ell}=m_0\left(\begin{array}{ccc} a & 1 & 1 \\ 1 & a & 1 \\
1 & 1 & a \end{array}\right)
\end{eqnarray}
where $m_0=h_1\kappa'_1$ and $m_0a=h_0\kappa'_0$. Three eigen vectors of 
this mass matrix can be written as:
\begin{eqnarray}
\left(\begin{array}{c} e\\ \mu \\ \tau \end{array}\right) =
\left(\begin{array}{ccc} 
\frac{1}{\sqrt{2}} & -\frac{1}{\sqrt{2}} & 0\\
\frac{1}{\sqrt{6}} &\frac{1}{\sqrt{6}} &-\frac{2}{\sqrt{6}} \\
\frac{1}{\sqrt{3}} &\frac{1}{\sqrt{3}} &\frac{1}{\sqrt{3}}
\end{array} \right)\left(\begin{array}{c} e^0 \\ \mu^0 \\ \tau^0\end{array} 
\right) 
\end{eqnarray}
where the particles in the above equation with superscript zero denote that
they are the weak eigenstates prior to the diagonalization of the mass 
matrix. Note that this matrix is precisely the matrix in Eq. (2). The 
corresponding eigenvalues are:
 \begin{eqnarray}
m_e=m_0(a-1)\nonumber \\
m_\mu=m_0(a-1)\nonumber\\
m_\tau=m_0(a+2)
\end{eqnarray}
It is then easy to see that if the Majorana mass matrix for the neutrinos 
is diagonal, the consequent neutrino mixing matrix determined by the 
diagonalization of the charged lepton matrix above and is precisely of 
type (C) described above. Note however that the muon and the electron 
masses are equal. In order for them to be much less than the $\tau$ mass
as observed, we must have $a\simeq 1$. It is interesting to note that if 
instead of $S_3$ symmetry, one assumes $S_{3L}\times S_{3R}$ then indeed 
one ends up with $a=1$ as has been noted already\cite{fritzsch}. As alluded 
to before, we must 
have permutation symmetry breaking terms to split their masses. Before 
proceeding to that discussion, let us turn to the neutrino sector to make 
sure that no mixing angles emerge in the neutrino sector that could vitiate 
the maximal pattern. Equation (3) and the $\phi_{\alpha}$ vev 
pattern imply that the Dirac mass matrix for the three 
neutrinos is diagonal with all $m_{{\nu^D}_\alpha}$ given by 
$m_0a(\kappa'_0/\kappa_0)$. In order to understand the small neutrino 
masses we must implement a seesaw mechanism. It turns out that in 
this case the appropriate one for us is the double seesaw mechanism 
discussed in \cite{valle}. From the terms
involving the gauge singlet fermions $\sigma_a$'s in the Lagrangian:
\begin{eqnarray}
{\cal L_{\sigma}} = \Sigma_a f (\bar{\psi}_{aR}\chi_{aR}\sigma_a ~+~ 
R\rightarrow L + m_{\sigma_a} \sigma^2_a) +~~ h. c.
\end{eqnarray}
the $(\nu_L,\nu_R,\sigma)$ mass matrix comes out to be
\begin{eqnarray}
M_{\nu}~=~\left(\begin{array}{ccc} 
0 & m_{\nu^D} & 0 \\
m_{\nu^D} & 0 & f v_R \\
0 & fv_R & M_{\sigma} \end{array} \right) 
\end{eqnarray}
with $<\chi_{aR}>= v_R$ providing the largest mass scale in the problem.
Each of the entries in Eq. (7) except the $M_\sigma$ is a $3\times 3$ unit 
matrix. In the limit of 
exact permutation symmetry, $M_{\sigma}$ would also be a unit matrix. The 
light neutrino eigenvalues are given by:
 \begin{eqnarray}
m_{\nu_a}\simeq \frac{m^2_{\nu^D}m_{\sigma_a}}{f^2 v^{2}_R}
\end{eqnarray}
It is important to emphasize that there is no mixing in the purely 
neutrino sector so that in the basis where the charged leptons are diagonal,
we have  the desired maximal mixing pattern.
This in our opinion is the big model building challenge that we have 
solved in this article. Clearly, if permutation symmetry had 
not been broken by the different $\sigma_a$ masses, the mixing matrix 
would have been arbitrary.

 To get a feeling for the scale of new physics $v_R$, we note that
$m_{\nu^D}\simeq (\kappa'_0/\kappa_0)(m_{\tau}/3)$ GeV. Therefore, assuming 
 $\kappa'_0/ \kappa_0 \sim 0.1$, we get $m_{\nu^D}\simeq 0.06$ GeV; 
and for $v_R= 10^5$ GeV and $f=2$, we get $m_{\nu_a}\simeq  0.9\times
10^{-4}(m_{\sigma_a}/GeV)$ eV.
If we choose $m_{\sigma_3}\simeq 500$ GeV and $m_{\sigma_{1,2}}\ll 
m_{\sigma_3}$, we get $m_{\nu_{\tau}}\simeq 4.5\times 10^{-2}$ eV, which 
is in the range required to solve the atmospheric neutrino puzzle. 

At this stage it might appear that the muon- and electron-neutrino masses 
can be chosen at will by adjusting the $m_{\sigma_{1,2}}$. But this is 
not so since the muon and electron masses which are tiny at the tree level 
(if we choose $a=1$) 
must also arise out of the mass splitting among the $\sigma_a$'s at the 
one loop level. The radiative contributions to the muon and electron 
masses arise from the 
diagram of type shown in Fig.1 and we can estimate this contribution to be:
\begin{eqnarray}
m^{(1)}_{\ell_a}\simeq 
\frac{f^2}{16\pi^2}\frac{m^2_{\sigma_a}\mu^3\kappa_0}{\lambda(\beta v_R)^5}
\end{eqnarray}
where $\beta v_R$ is the typical heavy Higgs boson mass that appears in 
the loop. We have also used the 
fact the vevs of $\chi_{a L,R}$ satisfy the relation $v_{aL} 
v_{aR}\simeq \frac{\kappa_0 \mu}{\lambda}$. 
\begin{figure}[htb]
\begin{center}
\epsfxsize=8.5cm
\epsfysize=8.5cm
\mbox{\hskip -1.0in}\epsfbox{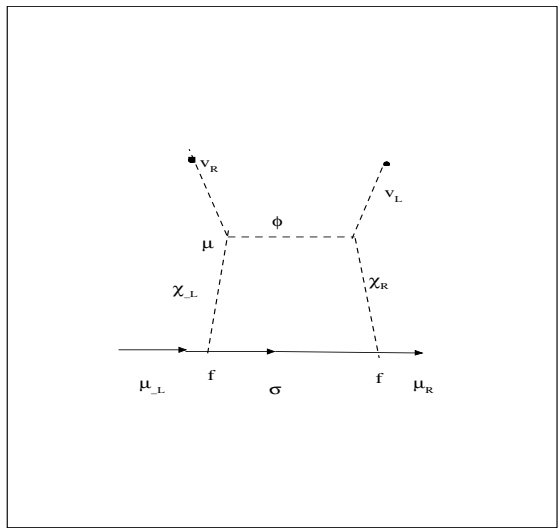}
\caption{ The Feynman diagram responsible for one loop radiative 
corrections to the muon and the electron masses. The dashed lines are the 
scalar bosons with appropriate quantum numbers.\label{Fig.1}} \end{center}  
  
\end{figure}
Choosing $\beta \simeq 0.14$ and $\mu\simeq v_R$, $\lambda\simeq 1$ and 
$f\simeq 2$, we estimate 
\begin{eqnarray}
 m^{(1)}_{\ell_a}\simeq 10^{-5} m^2_{\sigma_a}
\end{eqnarray}
Note that since we need to get the entire masses for the muon and the 
electron from the one loop correction, we must choose $m_{\sigma_2}\simeq 
100$ GeV
and $m_{\sigma_1}\simeq 7$ GeV. This then implies that $m_{\nu_{\mu}}\simeq
9\times 10^{-3}$ eV and $m_{\nu_e}\simeq 6\times 10^{-4}$ eV. We thus see 
that $\Delta m^2_{12}$ relevant for solving the solar neutrino problem is 
$\simeq 8.1\times 10^{-5}$ eV$^2$. This is comfortably in the right range 
for solving the solar neutrino problem using the large angle MSW solution.

\section{Higgs potential and symmetry breakings}

Let us now discuss the vev pattern assumed in the preceding analysis.
Two points need to be discussed are (i) the specific vev pattern for the 
field 
$\phi_{1,2}$ that differentiates the neutrino Dirac mass from the charged 
lepton mass matrix and (ii) the induced $\chi_{aL}$ vev. Note that due to 
the nontrivial transformation of the $\phi_1$ field under the $Z_3$ 
symmetry, the only allowed terms involving it in the potential are 
$Tr(\phi^{\dagger}_1\phi_1)$, 
$Tr(\phi^{\dagger}_1\phi_1\phi^{\dagger}_1\phi_1)$, 
$Tr(\phi^{\dagger}_1\phi_0\phi^{\dagger}_0\phi_1)$. Similar thing happens 
for $\phi_2$.
 Note further that the $Z_4$ symmetry forbids terms like $Tr 
\tilde{\phi}^{\dagger}_1\phi_2$. The absence of terms 
of the form $Tr \tilde{\phi}^{\dagger}_{1,2}\phi_{1,2}$ guarantees that 
once we 
choose the vev of the form $Diag<\phi_{1,2}>=(0,\kappa'_{1,2})$, there 
are no 
tadpole like terms that can destabilize that vacuum. Finally the fact 
that under left-right symmetry $\phi_1\leftrightarrow \phi^{\dagger}_2$
guarantees there is a discrete symmetry between these two fields leading 
to a stable minimum with $\kappa'_1=\kappa'_2$.

Turning now to the second point, note that the potential involving the 
$\chi_{aL,R}$ fields has the form
\begin{eqnarray}
V(\chi_{aL},\chi_{aR})= - 
M^2_0(\chi^{\dagger}_{aL}\chi_{aL}+\chi^{\dagger}_{aR}\chi_{aR}\nonumber\\
+\lambda_+(\chi^{\dagger}_{aL}\chi_{al}+
\chi^{\dagger}_{aR}\chi_{aR})^2 \nonumber\\
+\lambda_-(\chi^{\dagger}_{aL}\chi_{al}-
\chi^{\dagger}_{aR}\chi_{aR})^2 +\mu_a\chi^{\dagger}_{aL}\phi_0\chi_{aR}+h.c.
\end{eqnarray}
where sum over $a$ has been omitted for simplicity.
Minimizing this we get that $v_{aL}v_{aR}\simeq (\mu \kappa_0)/\lambda$.

A few comments about the model are in order.

\noindent (i) It is worth pointing out that in presence of the permutation 
symmetry breaking terms in the singlet fermion sector, there will be small 
deviations from the equality of the $\mu_a$'s and consequently of the scalar 
doublet masses. But these effects are small and they do not alter any of 
our conclusions.

\noindent (ii) The quark fields are assumed to be singlets under 
$S_3 \times Z_3\times Z_2$. Therefore, their masses arise from the $\phi_0$ 
couplings only and thus are not constrained by the patterns in the lepton 
sector.

\noindent(iii) The lightest of the singlet fermions $\sigma_1$ which 
couples to electrons can be produced at LEP energies but has a cross section
of order $\sigma_{e^+e^-\rightarrow \sigma_1\sigma_1}\sim f^4E^2/v^4_R$ 
which at the highest LEP energies is about $\sim 10^{-44} f^4$ cm$^2$ and
is thus practically invisible.

\section{Conclusion and outlook}
In conclusion, we have succeeded in constructing a natural gauge model 
for the democratic maximal mixing for neutrinos suggested by the present 
neutrino data if LSND results are not included. The models also predicts 
a small mass difference between the $\nu_e$ and $\nu_{\mu}$ as needed 
for the large angle MSW solution to the solar neutrino problem. To the 
best of our knowledge, this is the first time that a gauge model for 
understanding the democratic lepton mass matrix in an extension of the 
standard model has been constructed. The model is essentially an 
electroweak scale model with low scale for the right handed $W$'s and 
uses the double seesaw mechanism to generate small neutrino masses. Some 
of the new fermions of the model are light in the sense of collider 
physics. But their couplings to known particles are weak and thus there 
is no conflict with existing data.

This work is supported by the National Science 
Foundation under grant no. PHY-9802551 and also a grant from the 
Israel-US Binational Science Foundation.

\begin{table}[htb]
\begin{center}
\[
\begin{array}{|c||c||c||c|}
\hline
Fields &Z_4 & Z_3 & Z_2 \\
\hline
\psi_{aL}& {\bf 1}  & \omega^a & {\bf 1}\\
\psi_{aR} &-i & \omega^a & {\bf 1} \\
\phi_0 & i & {\bf 1} &{\bf 1} \\
\phi_1  & i & \omega^{-1} & {\bf 1} \\
\phi_2 & i & \omega^{-2} & {\bf 1} \\
\chi_{aR}&-i  & \omega^{a}& (-1)^a\\
\chi_{aL} & {\bf 1} & \omega^{a} & (-1)^a \\
\sigma_a  &{\bf 1} & {\bf 1} & (-1)^a \\
\hline
\end{array}
\] 
\end{center}
\caption{Transformation properties of the various fields under  
$Z_4\times Z_3\times Z_2$} \label{table}
\end{table}

\end{document}